\documentclass[english,aps,prd,onecolumn,superscriptaddress,nofootinbib,preprint]{revtex4-1}
\usepackage[T1]{fontenc}
\usepackage[latin9]{inputenc}

\usepackage{amsfonts}
\usepackage{amssymb}
\usepackage{amsmath}
\usepackage{amsthm}
\usepackage[english]{babel}
\usepackage{hyperref}
\usepackage{natbib}
\usepackage{marginnote}
\usepackage{color}

\makeatletter

\begin{document}

\title{ADM Analysis of Gravity Models within the Framework of Bimetric Variational Formalism}
\preprint{HIP-2014-31/TH}

\author{Alexey Golovnev}
\email{agolovnev@yandex.ru}
\affiliation{Saint-Petersburg State University, High Energy Physics Department\\Ulyanovskaya ul., d. 1; 198504 Saint-Petersburg, Petrodvoretz; Russia}

\author{Mindaugas Kar\v{c}iauskas}
\email{mindaugas.karciauskas@helsinki.fi}
\author{Hannu J. Nyrhinen}
\email{hannu.nyrhinen@helsinki.fi}

\affiliation{University of Helsinki and Helsinki Institute of Physics\\P.O. Box 64, FI-00014, Helsinki, Finland}

\begin{abstract}
Bimetric variational formalism was recently employed to construct novel bimetric gravity models. In these models an affine connection is generated by an additional tensor field which is independent of the physical metric. In this work we demonstrate how the ADM decomposition can be applied to study such models and provide some technical intermediate details. Using ADM decomposition we are able to prove that a linear model is unstable as has previously been indicated by perturbative analysis. Moreover, we show that it is also very difficult if not impossible to construct a non-linear model which is ghost-free within the framework of bimetric variational formalism. However, we demonstrate that viable models are possible along similar lines of thought. To this end, we consider a set up in which the affine connection is a variation of the Levi-Civita one. As a proof of principle we construct a gravity model with a massless scalar field obtained this way.
\end{abstract}

\maketitle

\section{Introduction}

A connection is a crucial quantity in gravity theories as it determines a parallel transport of vectors. In Einstein's formulation of gravity it is a unique function of the metric, which in its own right determines distances and angles. However, such a relation between the two is not required by any fundamental principle. Indeed, this arbitrariness is recognised in ''Palatini variation formalism'', in which connection is treated as an independent variable. 

If one is to introduce a new degree of freedom into a theory, one would expect it to be described in a coordinate independent way. That is, one expects it to be a tensor quantity. However, connection is not a tensor. Therefore one is led to consider a generalisation, in which connection is not a fundamental degree of freedom by itself but rather generated by an independent tensor field instead. Even if this tensor field does not participate in lengths and angles, it is generally also called ''a metric''. To make a distinction we call the tensor field that couples directly to matter ''a physical metric''. Hence the name ''Bimetric Variational Formalism''. One has to bear in mind that this is a bimetric incarnation of the metric-affine approach which is not to be confused with very popular bimetric gravity theories stemming from the ghost-free massive gravity models. 

Models, such as the so called {\it C}-theories \cite{AEK}, exist where the two metrics are connected by a particular functional relation. Building upon these theories  more general setups have also been studied. In ref. \cite{Tomi} the author considers a model in which the physical metric  ${\hat g}_{\mu\nu}$ determines the metric structure of a manifold and also appears in the measure factor and in the matter Lagrangian. The other metric $g_{\mu\nu}$ generates the connection, which is  used for constructing the Ricci tensor.

First, consider a model of this type with the action 
\begin{equation}
\label{initial}
S=\int d^4  x\sqrt{\hat g}~\mathcal R.
\end{equation}
Here $\mathcal R\equiv{\hat g}^{\mu\nu}R_{\mu\nu}$, where ${\hat g}_{\mu\nu}$ is the physical metric  and the Ricci tensor $R_{\mu\nu}$ comprises the Levi-Civita type connection of the second metric $g_{\mu\nu}$
\begin{equation}
\label{gamma}
\Gamma_{\mu\nu}^{\alpha}=\frac{1}{2}g^{\alpha\beta}\left(\partial_{\mu}g_{\beta\nu}+\partial_{\nu}g_{\mu\beta}-\partial_{\beta}g_{\mu\nu}\right).
\end{equation} 
Note that one might say that $\hat g$ is non-dynamical in this  model while $g$ contains higher derivatives and, consequently, the Ostrogradski ghosts. However, it is not that straightforward. We have to integrate those terms by parts. After that it is easy to see that both metrics are dynamical, and the velocities of $\hat g$ enter the action linearly. As we will discuss in detail in the next Section, the non-perturbative analysis shows that it is the latter fact which implies the presence of ghosts.

A perturbative analysis of the action in eq.  \eqref{initial}  shows that this model is indeed unhealthy. In ref. \cite{previous} the authors expand both metrics around Minkowski background. They find that at quadratic level the action (\ref{initial}) consists of two copies of the graviton action whose kinetic energies have opposite signs (the healthy part being the perturbation of ${\hat g}_{\mu\nu}$ and the unhealthy part being the difference of the two metrics). 
The authors  also conjecture  that a linear combination of the action \eqref{initial} and the action of general relativity might give a viable model by reversing the sign of the quadratic action of unhealthy modes.
However, the quadratic action only takes into account helicity 2 modes while being a strong coupling limit for other degrees of freedom.
The behaviour of the  latter can not be determined using only  a quadratic approximation, but nevertheless it can be dangerous.
In particular,  the conformal mode of the metric difference gives rise to a healthy scalar degree of freedom \cite{Tomi}. Thus, it would be natural to suspect that its kinetic term gets reversed to the negative sign by any combination that cures the helicity 2 modes.

In this paper we present a non-perturbative analysis of  bimetric variational formalism. In Section \ref{sec:Simple} we analyse the model in eq. \eqref{initial}. We  confirm that this model is unhealthy. 
In Section \ref{sec:GR} we extend our analysis to a model in which the action in eq. \eqref{initial} is linearly added to the action of general relativity. We find that it does not make the model more stable around the Minkowski solution.
In Section \ref{sec:f(r)} we show that also non-linear $f(\mathcal{R})$-type modifications 
are unhealthy. In Section \ref{secHealthy} we consider the connection \eqref{gamma} as a particular variation of the Levi-Civita connection of the physical metric ${\hat g}_{\mu\nu}$. We show that this class of models can be healthy and provide a simple example of such a model.
Finally, in Section 6 we conclude.

\section{The Simplest Form of Bimetric Variational Formalism\label{sec:simplest} \label{sec:Simple}}

To perform a non-perturbative Hamiltonian analysis, we write the action in eq. \eqref{initial} in terms of ADM variables defined as
\begin{equation}
\label{ADMhat}
ds^2 = g_{\mu\nu}dx^{\mu}dx^{\nu} \equiv -\left(N^2-N_i N^i \right)dt^2+2N_i dx^i dt+\gamma_{ij}dx^i dx^j.
\end{equation}
Here $N$ and $N_i$ are lapse and shift functions respectively, and it should be understood that the spatial indices are lowered and raised with the spatial metric $\gamma_{ij}$ and its inverse $\gamma^{ij}$. From the definition in eq. (\ref{ADMhat}) we can read the metric components $ g_{00}=-\left(N^2-N_i N^i \right)$, $g_{0i}=N_i$ and $g_{ij}=\gamma_{ij}$.
It is also straightforward to check that the inverse metric is given by
$g^{00}=-\frac{1}{N^2}$, $g^{0i}=\frac{N^i}{N^2}$ and $g^{ij}=\gamma^{ij}-\frac{N^i N^j}{N^2}$.
Comparing this result with the Cramer's rule for $g^{00}$ we get the useful relation for
determinants, $\sqrt{-g}=N\sqrt{\gamma}$.

Using these variables Riemann tensor components are given by \cite{ADM,book} 
\begin{eqnarray}
R_{ijkl}&=&{\mathop{R}\limits^{(3)}}\vphantom{R}_{ijkl}+K_{ik}K_{jl}-K_{il}K_{jk},\label{ijkl}\\
n_{\mu}{R^{\mu}}_{ijk}&=&{\mathop{\bigtriangledown}\limits^{(3)}}\vphantom{\Gamma}_j K_{ki} - {\mathop{\bigtriangledown}\limits^{(3)}}\vphantom{\Gamma}_k K_{ji},\label{0ijk}\\
n^{\mu}n^{\nu}R_{\mu i\nu j}&=&\frac1N\left(\dot{K}_{ij}+{\mathop{\bigtriangledown}\limits^{(3)}}\vphantom{\Gamma}_i {\mathop{\bigtriangledown}\limits^{(3)}}\vphantom{\Gamma}_j N+N{K_i}^k K_{kj}-{\mathcal Lie}_{\overrightarrow N}K_{ij}\right),\label{0i0j}
\end{eqnarray}
where $n^{\mu}$ is  a unit vector 
$n^{\mu}\equiv\left(\frac{1}{N}, -\frac{N^i}{N}\right)$ which is normal to the three-dimensional slices and corresponds to
the one-form $n_{\mu}\equiv\left(-N, {\overrightarrow 0}\right)$. The extrinsic curvature is given by
\begin{equation}
K_{ij}=\frac{1}{2N} \left({\mathop{\bigtriangledown}\limits^{(3)}}\vphantom{\Gamma}_i N_j+
{\mathop{\bigtriangledown}\limits^{(3)}}\vphantom{\Gamma}_j N_i - \dot{\gamma}_{ij}\right),\end{equation}
where the 3-dimensional covariant derivatives are taken with respect to $\gamma_{ij}$ and the Lie derivative is defined as
\begin{equation}{\mathcal Lie}_{\overrightarrow N}K_{ij}\equiv N^k\partial_k K_{ij}+K_{ik}\partial_j N^k+K_{jk}\partial_i N^k.\end{equation}
Note that the Lie derivative is a tensor and moreover, the ordinary partial derivatives can be substituted by the covariant ones. This does not change  the Lie derivative as long as the connection is symmetric.

Using eqs.~(\ref{ijkl})--(\ref{0i0j}) the Ricci tensor $R_{\mu\nu}\equiv R^{\alpha}_{\mu\alpha\nu}$ can be expressed in terms of standard ADM variables as
\begin{eqnarray}
R_{ij} &=& {\mathop{R}\limits^{({\mathit 3})}}_{ij}-\frac1N \dot K_{ij}-\frac1N{\mathop{\bigtriangledown}\limits^{({\mathit 3})}}_{i}{\mathop{\bigtriangledown}\limits^{({\mathit 3})}}_{j}N+K_{ij} K^k_k -2 K_{ik}K^k_j+\frac1N{\mathcal Lie}_{\overrightarrow N}K_{ij},\label{ij}\\
n^{\mu}R_{\mu i} &=& {\mathop{\bigtriangledown}\limits^{({\mathit 3})}}_{i}K^j_j-{\mathop{\bigtriangledown}\limits^{({\mathit 3})}}_{j}K_i^j,\label{0i}\\
n^{\mu}n^{\nu}R_{\mu\nu} &=& \frac1N \gamma^{ij}\dot K_{ij}+\frac1N{\mathop{\bigtriangleup}\limits^{({\mathit 3})}}N+K_{ij} K^{ij}-\frac1N\gamma^{ij}{\mathcal Lie}_{\overrightarrow N}K_{ij}.\label{00}
\end{eqnarray}

We can perform the same (3+1)-decomposition of the physical metric ${\hat g}_{\mu\nu}$  by changing the above ADM variables to, say, $M$, $M^i$ and $\hat\gamma_{ij}$. To find the scalar ${\hat g}^{\mu\nu}R_{\mu\nu}$, we then use eqs.~(\ref{ij})--(\ref{00}) to express components of the Ricci tensor as
\begin{eqnarray}
R_{0i}&=&Nn^{\mu}R_{\mu i}+N^j R_{ij},\\
R_{00} &=& N^2 n^{\mu}n^{\nu}R_{\mu\nu}+2NN^i n^{\mu}R_{\mu i}+N^i N^j R_{ij}.
\end{eqnarray}
Plugging this result into the action in eq.~\eqref{initial} the Lagrangian density of this model becomes
\begin{multline}
\label{ADMinitial}
\sqrt{- \hat g}{\hat g}^{\mu\nu}R_{\mu\nu}=M\sqrt{\hat \gamma}\left[-\frac{N^2}{M^2}\left(\frac1N\gamma^{ij}{\dot K}_{ij}+K_{ij}K^{ij}+\frac1N{\mathop{\bigtriangleup}\limits^{(\mathit 3)}}N-\frac1N\gamma^{ij}{\mathcal Lie}_{\overrightarrow N}K_{ij}\right) \right. \\
+2\frac{N}{M^2}(M^i-N^i)\left({\mathop{\bigtriangledown}\limits^{(\mathit 3)}}_iK^j_j -{\mathop{\bigtriangledown}\limits^{(\mathit 3)}}_j K^j_i\right)\\
+\left.\left(\hat\gamma^{ij}-\frac{\left(M^i-N^i\right)\left(M^j-N^j\right)}{M^2}\right) \left({\mathop{R}\limits^{(\mathit 3)}}_{ij}-\frac1N {\dot K}_{ij}-\frac1N{\mathop{\bigtriangledown}\limits^{(\mathit 3)}}_i{\mathop{\bigtriangledown}\limits^{(\mathit 3)}}_j N+K_{ij}K^k_k-2K_{ik}K^k_j+ \frac1N {\mathcal Lie}_{\overrightarrow N} K_{ij}\right)\right]
\end{multline}
where indices of extrinsic curvature tensors are raised by $\gamma^{ij}$.

Defining new variables $a\equiv\frac{N}{M}$ and $a^i\equiv\frac{M^i-N^i}{M}$ and collecting similar terms in eq.~(\ref{ADMinitial}) together, we get
\begin{multline}
\label{ADMinitial-a}
\sqrt{-\hat g}{\hat g}^{\mu\nu}R_{\mu\nu}=\sqrt{\gamma}\cdot\sqrt{\frac{\hat\gamma}{\gamma}}\left[\left(\frac1a\left(\hat\gamma^{ij}-a^i a^j\right)+a\gamma^{ij}\right)\left({\mathcal Lie}_{\overrightarrow N}K_{ij}-{\mathop{\bigtriangledown}\limits^{(\mathit 3)}}_i {\mathop{\bigtriangledown}\limits^{(\mathit 3)}}_j N-{\dot K}_{ij}\right) \right.\\
+2Na^i\left({\mathop{\bigtriangledown}\limits^{(\mathit 3)}}_i K^j_j - {\mathop{\bigtriangledown}\limits^{(\mathit 3)}}_j K^j_i\right) + \frac{N}{a}\left(\hat\gamma^{ij}-a^i a^j \right){\mathop{R}\limits^{(\mathit 3)}}_{ij}\\
+\left.N\left(\frac{1}{a}\left(\hat\gamma^{ij}-a^i a^j\right)\gamma^{kl} -a\gamma^{ik}\gamma^{jl}-\frac{2}{a}\left(\hat\gamma^{ik}-a^i a^k\right)\gamma^{jl}\right)K_{ij}K_{kl}\right],
\end{multline}
where we keep an explicit factor $\sqrt\gamma$ to simplify the  integration by parts of terms containing ${\mathop{\bigtriangledown}\limits^{(\mathit 3)}}_i $.

Now we  see that the second time derivative enter the action only in one place, namely $-\sqrt{\gamma}\sqrt{\frac{\hat\gamma}{\gamma}}\frac{1}{a}\left(\hat\gamma^{ij}-a^i a^j +a^2 \gamma^{ij}\right){\dot K}_{ij}$. Defining a metric combination $\chi^{ij}$ as
\begin{equation}
\label{chi}
\chi^{ij}\equiv\sqrt{\frac{\hat\gamma}{\gamma}}\frac{1}{a}\left(\hat\gamma^{ij}-a^i a^j +a^2 \gamma^{ij}\right)
\end{equation} 
the integration by parts gives
\begin{equation}
\label{byparts}
-\sqrt{\gamma}\chi^{ij}{\dot K}_{ij}\longrightarrow \sqrt{\gamma}K_{ij}{\dot \chi}^{ij}+\sqrt{\gamma}K_{ij}\chi^{ij}{\mathop{\bigtriangledown}\limits^{(\mathit 3)}}_k N^k-\sqrt{\gamma}N\chi^{ij}\gamma^{kl}K_{ij}K_{kl}.
\end{equation}
This makes $\chi^{ij}$ dynamical and its velocities enter the Lagrangian linearly.
Therefore, the kinetic function is unbounded on both sides. This is precisely the problem of ghosts. Indeed, the (quadratic in velocities) kinetic part is of the form ${\mathcal A}KK+{\dot\chi} K$. It is enough to fix $K$ to some value and vary  $\dot\chi$
 to see that it is not bounded from either side. More formally, if we find any eigenvector of the (6-dimensional) quadratic form ${\mathcal A}$, for example along the z-axis, then $AK^2_z+{\dot\chi}_z K_z=A\left(\left(K_z+\frac{1}{2A}{\dot\chi}_z\right)^2-\frac{1}{4A^2}{\dot\chi}^2_z\right)$ which is not positive  definite for either sign of  $A$.

Using eq.~\eqref{byparts} to integrate the action with the Lagrangian in eq.~\eqref{ADMinitial-a} by parts it can be rewritten as
\begin{multline}
\label{ADMinitial-chi}
S=\int dt d^3 x\sqrt{\gamma}\left[2\chi^{ij}K_{jk}{\mathop{\bigtriangledown}\limits^{(\mathit 3)}}_i N^k-K_{ij}N^k{\mathop{\bigtriangledown}\limits^{(\mathit 3)}}_k \chi^{ij}-\chi^{ij}{\mathop{\bigtriangledown}\limits^{(\mathit 3)}}_i {\mathop{\bigtriangledown}\limits^{(\mathit 3)}}_j N+K_{ij}{\dot\chi}^{ij} \right.\\
+2\gamma^{kj}K_{ij}{\mathop{\bigtriangledown}\limits^{(\mathit 3)}}_k\left(\sqrt{\frac{\hat\gamma}{\gamma}}Na^i\right) - 2\gamma^{ij}K_{ij}{\mathop{\bigtriangledown}\limits^{(\mathit 3)}}_k \left(\sqrt{\frac{\hat\gamma}{\gamma}}Na^k\right) + N\left(\chi^{ij}-a\sqrt{\frac{\hat\gamma}{\gamma}}\gamma^{ij} \right){\mathop{R}\limits^{(\mathit 3)}}_{ij}\\
+\left.N\left(\left(a\sqrt{\frac{\hat\gamma}{\gamma}}\gamma^{ik}-2\chi^{ik}\right)\gamma^{jl}-a\sqrt{\frac{\hat\gamma}{\gamma}}\gamma^{ij}\gamma^{kl}\right)K_{ij}K_{kl}\right],
\end{multline}
where $\sqrt{\hat\gamma}$ has to be understood as a non-linear function of $\gamma_{ij}$, $\chi^{ij}$, $a$ and $a^i$. 
We also integrated terms with the three-derivative ${\mathop{\bigtriangledown}\limits^{(\mathit 3)}}$ by parts to dispose of the derivatives of $K_{ij}$ in preparation for the next section, where canonical momenta are computed.
Note also that we have combined the $-\sqrt{\gamma}K_{ij}{\mathop{\bigtriangledown}\limits^{(\mathit 3)}}_k \left(N^k \chi^{ij}\right)$ term from the spatial integration by parts in one of the Lie derivative terms with the $\sqrt{\gamma}K_{ij}\chi^{ij}{\mathop{\bigtriangledown}\limits^{(\mathit 3)}}_k N^k$ term from the temporal integration by parts in $-\sqrt{\gamma}\chi^{ij}{\dot K}_{ij}$. This produces the $-\sqrt{\gamma}K_{ij}N^k{\mathop{\bigtriangledown}\limits^{(\mathit 3)}}_k \chi^{ij}$ term in the action in eq. \eqref{ADMinitial-chi} above .

Although $a$ and $a^i$  acquire time derivatives, they enter the action only within the ${\dot\chi}^{ij}$ term. Therefore they are non-dynamical.
However, without derivatives they enter the $\sqrt{\hat\gamma}$ term non-linearly. Indeed, $\frac{\hat\gamma}{\gamma}={\rm det}\left({\hat\gamma}^{ik}\gamma_{kj}\right)$ and from the definition in eq.~\eqref{chi} we see that $\sqrt{\frac{\hat\gamma}{\gamma}}$ is a solution of a non-linear algebraic equation
\begin{equation} \frac{\hat\gamma}{\gamma}={\rm det}\left(\sqrt{\frac{\gamma}{\hat\gamma}}a\chi^i_j+a^i a_j-a^2\delta^i_j\right), \end{equation}
where indices are raised by $\gamma^{ij}$.
It means that, while the variations with respect to $N$ and $N^i$ produce the physical constraints, the other non-dynamical variables $a$ and $a^i$ give only equations for themselves. Therefore, we have 12 variables with time derivatives (6 for $\gamma_{ij}$ and 6 for $\chi^{ij}$) in the action and 4 gauge invariances. Thus, the total number of degrees of freedom is 8.
It should not come as a surprise since this is what is generically obtained from bigravity models unless a special care is taken to get rid off the Boulware-Deser ghost.

\subsection{The Hamiltonian}

For completeness, let us obtain the Hamiltonian of the model. We denote the properly symmetrised version of the quadratic form \begin{equation}
A^{ijkl} \equiv \left(a\sqrt{\frac{\hat\gamma}{\gamma}}\gamma^{ik}-2\chi^{ik}\right)\gamma^{jl}-a\sqrt{\frac{\hat\gamma}{\gamma}}\gamma^{ij}\gamma^{kl},
\end{equation}
and compute the canonical momenta
\begin{equation}
\label{momentum-chi}
\pi^{(\chi)}_{ij}\equiv\frac{\partial\mathcal L}{\partial{\dot\chi}^{ij}}=\sqrt{\gamma}K_{ij}=\frac{\sqrt{\gamma}}{2N}\left({\mathop{\bigtriangledown}\limits^{(\mathit 3)}}_i N_j+{\mathop{\bigtriangledown}\limits^{(\mathit 3)}}_j N_i -{\dot\gamma}_{ij}\right)
\end{equation}
and
\begin{multline}
\label{momentum-gamma}
\pi_{(\gamma)}^{ij}\equiv\frac{\partial\mathcal L}{\partial{\dot\gamma}_{ij}}=-\frac{\sqrt{\gamma}}{2N}\left[\chi^{ik}{\mathop{\bigtriangledown}\limits^{(\mathit 3)}}_k N^j+\chi^{jk}{\mathop{\bigtriangledown}\limits^{(\mathit 3)}}_k N^i-N^k{\mathop{\bigtriangledown}\limits^{(\mathit 3)}}_k \chi^{ij}+{\dot\chi}^{ij} \right.\\
+\left. \gamma^{kj}{\mathop{\bigtriangledown}\limits^{(\mathit 3)}}_k\left(\sqrt{\frac{\hat\gamma}{\gamma}}Na^i\right)+\gamma^{ki}{\mathop{\bigtriangledown}\limits^{(\mathit 3)}}_k\left(\sqrt{\frac{\hat\gamma}{\gamma}}Na^j\right) -2\gamma^{ij}{\mathop{\bigtriangledown}\limits^{(\mathit 3)}}_k \left(\sqrt{\frac{\hat\gamma}{\gamma}}Na^k\right) +N\left(A^{ijkl}+A^{klij}\right)K_{kl}\right].
\end{multline}
From equation~\eqref{momentum-chi} we can find the time derivatives of $\gamma$ as
\begin{equation}{\dot\gamma}_{ij}={\mathop{\bigtriangledown}\limits^{(\mathit 3)}}_i N_j+{\mathop{\bigtriangledown}\limits^{(\mathit 3)}}_j N_i-\frac{2N}{\sqrt{\gamma}}\pi^{(\chi)}_{ij}\end{equation}
or $K_{ij}=\frac{1}{\sqrt{\gamma}}\pi^{(\chi)}_{ij}$. The same can be done for $\dot\chi$ using eq. (\ref{momentum-gamma}), but we would not write this expression explicitly because the $\pi^{(\chi)}_{ij}{\dot\chi}^{ij}$ term cancels out with the $\sqrt{\gamma}K_{ij}{\dot\chi}^{ij}$ term in the Lagrangian when computing the Hamiltonian.

Finally, we can write the Hamiltonian as
\begin{multline}
\label{Hamiltonian}
H = \pi_{(\gamma)}^{ij}{\dot\gamma}_{ij}+\pi^{(\chi)}_{ij}{\dot\chi}^{ij}-{\mathcal L} \\
=-\int dt d^3 x\sqrt{\gamma}\left\{N\left[2\frac{\pi_{(\gamma)}^{ij}\pi^{(\chi)}_{ij}}{\gamma}+\left(\left(a\sqrt{\frac{\hat\gamma}{\gamma}}\gamma^{ik}-2\chi^{ik}\right)\gamma^{jl}-a\sqrt{\frac{\hat\gamma}{\gamma}}\gamma^{ij}\gamma^{kl}\right)\frac{\pi^{(\chi)}_{ij}\pi^{(\chi)}_{kl}}{\gamma} \right.\right. \\
 +\left.\left(\chi^{ij}-a\sqrt{\frac{\hat\gamma}{\gamma}}\gamma^{ij} \right){\mathop{R}\limits^{(\mathit 3)}}_{ij}+2\sqrt{\frac{\hat\gamma}{\gamma}}a^i\gamma^{kj}\left({\mathop{\bigtriangledown}\limits^{(\mathit 3)}}_i\frac{\pi^{(\chi)}_{kj}}{\sqrt{\gamma}} - {\mathop{\bigtriangledown}\limits^{(\mathit 3)}}_j \frac{\pi^{(\chi)}_{ki}}{\sqrt{\gamma}}\right) -{\mathop{\bigtriangledown}\limits^{(\mathit 3)}}_i{\mathop{\bigtriangledown}\limits^{(\mathit 3)}}_j\chi^{ij}\right]\\
 +\left.N^k\left(2\gamma_{kj}{\mathop{\bigtriangledown}\limits^{(\mathit 3)}}_i\frac{\pi_{(\gamma)}^{ij}}{\sqrt{\gamma}}-2{\mathop{\bigtriangledown}\limits^{(\mathit 3)}}_i\left(\chi^{ij}\frac{\pi^{(\chi)}_{jk}}{\sqrt{\gamma}}\right)-\frac{\pi^{(\chi)}_{ij}}{\sqrt{\gamma}}{\mathop{\bigtriangledown}\limits^{(\mathit 3)}}_k\chi^{ij}\right)\right\}.
\end{multline}
Again  we integrated some terms by parts  to express $N$ and $N^k$ as Lagrange multipliers, without (spatial) derivatives. We see from eq. (\ref{Hamiltonian}) that the Hamiltonian of a time-reparametrisation-invariant model is a combination of constraints, as it should be. Of course, perturbing around a fixed background which possesses  time translation invariance, a meaningful notion of perturbation energy can be restored. In that case, this energy is not positive definite because the $\gamma$-momenta $\pi_{(\gamma)}^{ij}$ enter the Hamiltonian in eq.~(\ref{Hamiltonian}) only linearly via the scalar term $\pi_{(\gamma)}^{ij}\pi^{(\chi)}_{ij}$.

\section{Combination with standard GR \label{sec:GR}}

In reference \cite{previous} an attempt is made to solve the ghost problem by combining the action in eq. (\ref{initial}) with the standard Einstein-Hilbert term for the  physical metric $\hat g$,
$$S=\int d^4  x\sqrt{\hat g}{\hat g}^{\mu\nu}\left(\alpha{\hat  R_{\mu\nu}}+ R_{\mu\nu}\right).$$
Contrary to intuition drawn from the quadratic action, this fails even worse than just reversing the sign of kinetic energy of the conformal mode. Indeed, in this case we have the Lagrangian density
\begin{multline}
\label{ADMcombined}
\sqrt{-\hat g}{\hat g}^{\mu\nu}\left(\alpha{\hat R}_{\mu\nu}+R_{\mu\nu}\right)=\alpha\sqrt{\hat\gamma}\left({\mathop{\hat R}\limits^{(\mathit 3)}}+\left({\hat\gamma}^{ik}{\hat\gamma}^{jl}-{\hat\gamma}^{ij}{\hat\gamma}^{kl}\right){\hat K}_{ij}{\hat K}_{kl}\right)\\
+\sqrt{\gamma}\cdot\sqrt{\frac{\hat\gamma}{\gamma}}\left[\left(\frac1a\left(\hat\gamma^{ij}-a^i a^j\right)+a\gamma^{ij}\right)\left({\mathcal Lie}_{\overrightarrow N}K_{ij}-{\mathop{\bigtriangledown}\limits^{(\mathit 3)}}_i {\mathop{\bigtriangledown}\limits^{(\mathit 3)}}_j N-{\dot K}_{ij}\right) \right.\\
+2Na^i\left({\mathop{\bigtriangledown}\limits^{(\mathit 3)}}_i K^j_j - {\mathop{\bigtriangledown}\limits^{(\mathit 3)}}_j K^j_i\right) + \frac{N}{a}\left(\hat\gamma^{ij}-a^i a^j \right){\mathop{R}\limits^{(\mathit 3)}}_{ij}\\
+\left.N\left(\frac{1}{a}\left(\hat\gamma^{ij}-a^i a^j\right)\gamma^{kl} -a\gamma^{ik}\gamma^{jl}-\frac{2}{a}\left(\hat\gamma^{ik}-a^i a^k\right)\gamma^{jl}\right)K_{ij}K_{kl}\right].
\end{multline}
An important difference from the action in eq. (\ref{ADMinitial-a}) is that we get independent kinetic terms for $\hat\gamma_{ij}$. Hence, we cannot absorb time derivatives of $a$ and $a^i$ into the $\chi$-field. 
After integrating the action with the above Lagrangian by parts, as detailed in eq.~(\ref{byparts}), $a$ and $a^i$ become 4 new degrees of freedom and their velocities enter the Lagrangian linearly. 
Also note that they are not pure gauge terms since, for example, solutions with ${\hat g}_{\mu\nu}=g_{\mu\nu}$ satisfy $a=1$ and $a^i=0$ for any choice of coordinates. And, after taking the gauge freedom into account, we can compute the number of degrees of freedom to be $6+6+4-4=12$. 

The temporal integration by parts and the subsequent determination of the canonical momenta are straightforward. However, the latter will mix different velocities, for example due to ${\hat K}K$-terms. Therefore expressing the velocities in terms of momenta and computing the Hamiltonian would be cumbersome, and we omit that.

Note that in ref. \cite{previous} an equivalent expression was given for the action in eq. (\ref{initial}). In our current notations, and summing up the infinite series $g^{\mu\nu}=
\left[\sum_{n=0}^{\infty}\left(-{\hat g}^{-1}h\right)^n\right]^{\mu}_{\alpha}{\hat g}^{\alpha\nu}$, it reads
\begin{multline}
\label{part-h-2}
S=\int d^4 x \sqrt{-\hat g}\left\{{\hat R}+
\frac14 
g^{\alpha\rho}
g^{\beta\sigma}
\left[\left({\hat\bigtriangledown}_{\beta}h_{\alpha\rho}\right){\hat g}^{\mu\kappa}
\left(2{\hat\bigtriangledown}_{\mu}h_{\sigma\kappa}-{\hat\bigtriangledown}_{\sigma}h_{\mu\kappa}\right)\right.\right.\\
-\left.\left.\left({\hat\bigtriangledown}_{\mu}h_{\beta\rho}+
{\hat\bigtriangledown}_{\beta}h_{\mu\rho}-{\hat\bigtriangledown}_{\rho}h_{\mu\beta}\right){\hat g}^{\mu\nu}
\left({\hat\bigtriangledown}_{\nu}h_{\alpha\sigma}+
{\hat\bigtriangledown}_{\alpha}h_{\nu\sigma}-{\hat\bigtriangledown}_{\sigma}h_{\alpha\nu}\right)\right]\right\}
\end{multline}
where $h_{\mu\nu}\equiv g_{\mu\nu}-{\hat g}_{\mu\nu}$ although we can safely substitute $g_{\mu\nu}$ for $h_{\mu\nu}$ because ${\hat g}_{\mu\nu}$ is covariantly constant with respect to ${\hat\bigtriangledown}_{\alpha}$.

We immediately see that the terms with ${\dot h}_{00}^2$, ${\dot h}_{00}{\dot h}_{0i}$ and ${\dot h}_{0i}{\dot h}_{0j}$ drop out from the action, but ${\dot h}_{00}{\dot h}_{ij}$ and ${\dot h}_{0k}{\dot h}_{ij}$ (and of course ${\dot h}_{ij}{\dot h}_{kl}$) do generically have non-zero coefficients. A straightforward computation yields only two non-zero terms in the kinetic function of $h_{00}$ and $h_{0i}$:  
\begin{eqnarray*}
\frac14\left(2g^{00}g^{0i}{\hat g}^{0j}+g^{00}g^{ij}{\hat g}^{00}-g^{00}g^{00}{\hat g}^{ij}-2g^{0i}g^{0j}{\hat g}^{00}\right){\dot h}_{00}{\dot h}_{ij},\\
\frac14\left(4g^{00}g^{jk}{\hat g}^{0i}+2g^{00}g^{0k}{\hat g}^{ij}-2g^{0k}g^{ij}{\hat g}^{00}-4g^{0i}g^{jk}{\hat g}^{00}\right){\dot h}_{0k}{\dot h}_{ij}.\end{eqnarray*}
When ${\hat g}_{\mu\nu}=g_{\mu\nu}$, these kinetic terms disappear in accordance with the linear analysis around the double-Minkowski solution which shows only helicity 2 modes. 
On the other hand, it nicely corresponds to the ADM formulation, where $\dot a$ and ${\dot a}^i$ enter the action linearly. 

The combination with pure Einstein-Hilbert term amounts to a mere change of the relative coefficient of the Einstein-Hilbert and the $\left({\hat\bigtriangledown}h\right)^2$-terms in the action in eq.~(\ref{part-h-2}). Surprisingly, this seemingly innocent deformation changes the number of degrees of freedom. However, one has to take into account that the pure GR part in this formulation is just an illusion. The covariant derivatives of $h_{\mu\nu}$ contain all the connection coefficients, and in particular the time derivatives of all ${\hat g}$-components. It yields a heavy kinetic mixing of ${\hat g}_{\mu\nu}$ and $g_{\mu\nu}$ (or $h_{\mu\nu}$) in the action in eq.~(\ref{part-h-2}). Therefore, counting the degrees of freedom is anything but an easy task in this formulation.

\section{Generalisations with non-linear functions of scalar invariants \label{sec:f(r)}}

In section~\ref{sec:simplest} it was  shown that the action in  eq.~\eqref{initial} has ghost degrees of freedom. However, one might still hope that applying the bimetric variational formalism to a non-linear generalisation of that action might produce a ghost free theory. In particular, consider an action of the form
\begin{equation}
\label{fRcurly}
S=\int d^4x\sqrt{-\hat g}~f\left(\mathcal R\right),
\end{equation}
where $f$ is some non-linear function of the scalar $\mathcal R\equiv{\hat g}^{\mu\nu}R_{\mu\nu}$.

Unfortunately, it turns out that none of the models in this class are ghost free. To prove this it is useful to rewrite the action in eq. \eqref{fRcurly} in an equivalent form, given by
\begin{equation}
\label{nl-trick}
S=\int d^4x\sqrt{-\hat g} \left[{\hat g}^{\mu\nu}f^{\prime}(\phi)R_{\mu\nu}+f(\phi)-\phi f^{\prime}(\phi) \right],
\end{equation}
where the prime denotes differentiation with respect to the scalar field $\phi$.
The equivalence of this form can be easily checked by finding the extremum of this action with respect to the field $\phi$ and plugging the resulting constraint back into the action. Such a transformation is often employed in conventional $f\left(R\right)$ models, where $R$ is the Ricci scalar constructed solely of the physical metric.

The $f^{\prime}(\phi)$ factor in eq. \eqref{nl-trick} can be absorbed by redefining the metric $\hat{g}_{\mu\nu}$. However, in contrast to  conventional $f(R)$ type models, the field $\phi$ does not become dynamical, because the Ricci tensor $R_{\mu\nu}$ depends on another metric. In particular, the action in eq. \eqref{nl-trick} is linear with respect to the factor ${\hat g}^{\mu\nu}R_{\mu\nu}$. To see this let us define the metric ${\hat{\tilde g}}_{\mu\nu}\equiv f^{\prime}(\phi)\cdot {\hat  g}_{\mu\nu}$. Plugging it into the action gives
\begin{equation}
\label{rescaled}
S=\int d^4x\sqrt{-\hat{\tilde g}}\left({\hat{\tilde g}}^{\mu\nu}R_{\mu\nu}+\frac{f(\phi)-\phi f^{\prime}(\phi)}{{f^{\prime}}^2(\phi)}\right).
\end{equation}
The first term of the result above is equivalent to the action in eq.~\eqref{initial}, which, as we have already shown, contains ghost degrees of freedom. Hence, $f(\mathcal R)$ type models are also unstable. 

The second term of the action in eq.~\eqref{rescaled} does not change this conclusion, as it is a constant. Its value is given at the extremum of the potential $V(\phi)=\left[f(\phi)-\phi f^{\prime}(\phi)\right]/{f^{\prime}}^2(\phi)$. Such a constant term can be interpreted as a cosmological constant. The aforementioned extremum exists if the condition
\begin{equation}
\label{extremum}
\phi f^{\prime}=2f
\end{equation}
is satisfied. 

As a side remark we note that when the function $f(\phi)$ is such that equation $f(0)=0$ is true, it is always possible to choose a solution $\phi=0$. In that case, the cosmological constant is zero and the above action is completely equivalent to the one in eq.~\eqref{initial}. Moreover, this is the only possible solution if $f(\phi)$ has the simplest dependence on $\phi$, namely $f(\phi)=\phi+\alpha \phi^2$. We also note that matter fields, which are omitted from the action in eq.~\eqref{fRcurly}, couple to the metric ${\hat g}_{\alpha\beta}=\frac{{\hat{\tilde g}}_{\alpha\beta}}{ f^{\prime}(\phi)}$. In effect, this renormalises Newton's constant.

We can also arrive at the same conclusion about the instability of the action in eq.~\eqref{fRcurly} by looking at equations of motion in ref.~\cite{Tomi}, which are computed directly from that action. The analogue of Einstein equations are obtained by varying that action with respect to ${\hat g}_{\mu\nu}$. We get
\begin{equation}
\label{eqT1}
f^{\prime}R_{\mu\nu} - \frac{1}{2}f {\hat g}_{\mu\nu} = 0.
\end{equation}
On the other hand, varying the same action with respect to $g_{\mu\nu}$ gives the equation for the metric ${\hat g}_{\mu\nu}$, which can be written as
\begin{equation}
\label{eqT2}
\left(g^{\mu\nu}\delta^\rho_\alpha\delta^\gamma_\beta
+ g^{\rho\gamma}\delta^{\mu}_\alpha\delta^\nu_\beta
  -   g^{\rho\nu}\delta^{\mu}_\alpha\delta^{\gamma}_\beta -  g^{\rho\nu}\delta^{\mu}_\beta\delta^{\gamma}_\alpha
\right) \bigtriangledown_{\gamma} \bigtriangledown_\rho \cdot \sqrt{\frac{\hat g}{g}} f^{\prime}{\hat g}^{\alpha\beta}=0.
\end{equation}

It is easy to see that in equation \eqref{eqT2} the factor of $f^{\prime}$ can be absorbed into redefinition of the physical metric as  ${\hat g}_{\alpha\beta}=\frac{{\hat{\tilde g}}_{\alpha\beta}}{ f^{\prime}}$. Now, the rescaled metric $ {\hat{\tilde g}}_{\alpha\beta}$ solves the same equation for the  linear model with $f^{\prime}=1$. This is also true for eq.~(\ref{eqT1}) if we take
$f({\mathcal R})={\mathcal R}\equiv {\hat g}^{\mu\nu}R_{\mu\nu}$ and a cosmological constant term with ${\hat\Lambda}=\frac{f-{\mathcal R}f^{\prime}}{{f^{\prime}}^2}$ is added. Indeed, dividing  eq. (\ref{eqT1}) by $f^{\prime}$, we get $R_{\mu\nu}-\frac12(\tilde{\mathcal{R}}+\hat\Lambda)\tilde{\hat g}_{\mu\nu}=0$. Finally, the condition in eq.~(\ref{extremum})  with $\phi=\mathcal{R}$ can be obtained by multiplying eq.~\eqref{eqT1} by ${\hat g}^{\mu\nu}$ and taking a trace. This shows that at the level of equations of motion actions \eqref{fRcurly} and \eqref{rescaled} are equivalent.

To summarise, to see whether ghost degrees of freedom present in the action in eq. \eqref{initial} could be avoided, we study its simplest non-linear generalisation \eqref{fRcurly}. We find that this generalisation fails to substantially modify the gravitational sector. Hence, we conclude that the action in eq.~\eqref{fRcurly} is as unstable as the original one and that the ghost cannot be avoided by a non-linear generalisation of the action in eq. \eqref{initial}.

One could still entertain other possible modifications in a hope to built a stable model. For example, one could envisage a model in which the action in eq.~\eqref{fRcurly} is augmented with additional standard Einstein-Hilbert term, the latter being constructed solely of the physical metric ${\hat g}_{\mu\nu}$. This procedure makes the field $\phi$ dynamical and non-minimally coupled to $\hat R$, after the metric in Einstein-Hilbert part is rescaled. As a result the first line in the action in eq.~\eqref{ADMcombined} is modified, but the modes which cause the instability, $a$ and $a^i$, stay unchanged.

Yet another hopeful modification of eq.~\eqref{fRcurly} could be an addition of a term with a scalar invariant ${\hat g}^{\mu\alpha}{\hat g}^{\nu\beta}R_{\mu\nu}R_{\alpha\beta}$. Unfortunately, such a construction is clearly unstable. This addition introduces terms that are quadratic in ${\dot K}$, which means that the action contains higher derivative terms and hence ghost degrees of freedom. 

\section{A Class of Healthy Models\label{secHealthy}}

The main aim of this work is to examine the stability of gravity theories in which the connection is constructed solely of a tensor field which is different from the physical metric. Such an approach can be called a bimetric variational formalism. By performing a non-perturbative analysis in sections \ref{sec:simplest} and \ref{sec:GR} we conclusively showed that models which are linear in $\mathcal R$ have ghost degrees of freedom. Faced with this result, in section \ref{sec:f(r)} we looked for non-linear models which adhere to the bimetric variational formalism and are healthy. Unfortunately, we discovered that it is very difficult to built such a model if not impossible.

In the last section of the paper we depart from the main goal of this work and investigate directions to construct models which do not follow strictly the philosophy of bimetric variational formalism. Namely, we relax the requirement for the affine connection to be generated solely of a tensor field which is different from the metric. Instead, we consider a case in which the affine connection is a modification of the Levi-Civita connection of the form 

\begin{equation}
\label{varG}
\Gamma^{\alpha}_{\mu\nu}={\hat\Gamma}^{\alpha}_{\mu\nu}+\delta \Gamma^{\alpha}_{\mu\nu},
\end{equation} 
where ${\hat\Gamma}^{\alpha}_{\mu\nu}$ is the Levi-Civita part generated by the physical metric ${\hat g}_{\mu\nu}$ and the second term is generated by additional degrees of freedom. In this case we have
\begin{equation}
\label{varRicci}
R_{\mu\nu}={\hat R}_{\mu\nu}+
{\hat\nabla}_{\alpha}\delta\Gamma^{\alpha}_{\mu\nu}-
{\hat\nabla}_{\nu}\delta\Gamma^{\alpha}_{\mu\alpha}+
\delta\Gamma^{\alpha}_{\beta\alpha}\delta\Gamma^{\beta}_{\mu\nu}-
\delta\Gamma^{\alpha}_{\mu\beta}\delta\Gamma^{\beta}_{\nu\alpha}.
\end{equation}
We can reasonably hope that some restricted class of such variations might give a healthy model. For example, consider $\delta\Gamma^{\beta}_{\mu\nu}$ to be a function of a scalar field and its gradients $X\equiv(\partial_\mu\rho)(\partial^\mu\rho)$ such that
\begin{equation}
\label{sG}
\delta\Gamma^{\alpha}_{\mu\nu}=\frac12 \left[a(\rho,X)\left(\delta^{\alpha}_{\mu}\partial_{\nu}\rho+\delta^{\alpha}_{\nu}\partial_{\mu}\rho\right)-b(\rho,X){\hat g}_{\mu\nu}\partial^{\alpha}\rho\right].
\end{equation}
Using eq.~\eqref{varRicci} we can rewrite the action in eq.~\eqref{initial} with the above  variation of connection as
\begin{equation}
\label{healthyPhi}
S=\int d^4\sqrt{-\hat g}\left[{\hat R}+\frac{3a^2-12ab+3b^2}{4}{\hat g}^{\mu\nu}(\partial_{\mu}\rho)(\partial_{\nu}\rho)\right].
\end{equation}

The case  $a=b=1$ is studied in ref.~\cite{Tomi}. In this work a conformal (Weyl) transformation is used to show that a model with this type of a connection is equivalent to general relativity with a healthy scalar field $\rho$. More generally, when $a$ and $b$ are some constants, we can absorb one of them into the definition of the scalar field. As can be seen from eq.~\eqref{healthyPhi}, the appearance of these constants is symmetric, and therefore we can arbitrarily choose either one of them without a loss of generality. Effectively this can be achieved by setting either of the constants to unity. For example, setting $b=1$, the condition for a correct sign of the kinetic term in eq.~\eqref{healthyPhi} reads\footnote{Had we chosen to absorb $a$ into the definition of the scalar field, effectively setting $a=1$, the same constraints would have to be satisfied by the constant $b$. More generally, one can make the symmetry between $a$ and $b$ explicit by writing an equivalent condition to eq.~\eqref{cond} as $(a-b)^2<2ab$.} 
\begin{equation}
\label{cond}
2-\sqrt3<a<2+\sqrt3.
\end{equation} 
Then it is clear from eq. \eqref{healthyPhi} that also in this case we reproduce Einstein's gravity with a propagating massless scalar field. 

Plugging the general case in eq.~\eqref{sG}, into eq.~\eqref{initial} we obtain a k-essence type of action with the scalar field Lagrangian given by ${\mathcal L}=f(\rho,X)\cdot X$. It remains to be seen how restrictive the factorisation of  $X$ turns out to be for potential applications. However, actions of this type have been used in the past to construct viable models of inflation \cite{kinflation,kInflPrtb} or dark energy, which avoid coincidence problems \cite{chiba,kessence}. The energy-momentum tensor of such a field can be written in a form of a perfect fluid with the pressure given by $p=\mathcal{L}$ and the energy density given by $\varepsilon = 2 X p_{,X}-p$  \cite{kinflation}. To ensure classical stability of the scalar field, the speed of sound $c_s^2\equiv p_{,X}/\varepsilon_{,X}$ must be positive $c_s^2>0$ \cite{kinflation,kHaloes}.

Let us generalise the action even further by considering it to be a non-linear function of $\mathcal{R}$ as in eq.~\eqref{fRcurly} with connection varied by $\delta\Gamma^{\alpha}_{\mu\nu}$ in eq.~(\ref{healthyPhi}) and $a=b=1$. Substituting ${\hat g}^{\mu\nu}R_{\mu\nu}={\hat g}^{\mu\nu}\left({\hat R}_{\mu\nu}-\frac32(\partial_{\mu}\rho)(\partial_{\nu}\rho)\right)$ into the action in eq.~(\ref{nl-trick}) and performing the metric rescaling we get an analogous result as in eq. (\ref{rescaled}), namely
\begin{equation*}
S=\int d^4x\sqrt{-\hat{\tilde g}}\left({\hat{\tilde R}}-\frac32(\partial_{\mu}\rho)(\partial^{\mu}\rho)-\frac32\left(\partial_{\mu}\log f^{\prime}(\phi)\right)\left(\partial^{\mu}\log f^{\prime}(\phi)\right)+\frac{f(\phi)-\phi f^{\prime}(\phi)}{{f^{\prime}}^2(\phi)}\right),
\end{equation*}
with matter coupling to ${\hat g}_{\mu\nu}=\frac{{\hat{\tilde g}}_{\mu\nu}}{ f^{\prime}(\phi)}$. Apart from the hidden massless scalar $\rho$, we see that the $f({\mathcal R})$ models with this type of modification are fully analogous to the standard $f(R)$ model.

The idea of connection being composite of the Levi-Civita part plus a non-metric part as in eq. \eqref{varG} is not new. Indeed, motivated  by constructing a theory which respects a local conformal invariance, Weyl considered a $\delta \Gamma^{\alpha}_{\mu\nu}$ of the form \cite{weyl}
\begin{equation}
\label{weylG}
\delta \Gamma^{\alpha}_{\mu\nu} = g^{\mu\sigma}\left(g_{\lambda\sigma}A_\nu+g_{\sigma\nu}A_\lambda-g_{\nu\lambda}A_\sigma\right),
\end{equation}
where $A_\mu$ is a vector field. If a gravitational action contains non-linear curvature terms, such a vector field becomes dynamical. This is the case in the theory studied by Weyl. As is well known, unfortunately equations of motion derived from an action with higher order curvature terms contain higher order derivatives and therefore also ghost degrees of freedom (see e.g. \cite{ostro} and references therein). 

Apart from the issue of local conformal invariance, the metric variation of the above type is studied in ref.~\cite{JT}. To avoid higher order equations of motion for the metric, the authors consider an action with curvature terms up to a quadratic order. This action reduces to the Gauss-Bonnet form when $\delta \Gamma^{\alpha}_{\mu\nu}$ in eq. \eqref{weylG} vanishes. They show that in four dimensional theory, the field $A_\mu$ satisfies the Proca equation of a massive vector field.

\section{Conclusions}

This paper describes basic techniques of performing the ADM analysis for models constructed with Bimetric Variational Formalism of ref. \cite{Tomi}. In these models connection is not generated by a physical metric, which determines distances and angles. Instead, it is generated by a completely independent tensor field, which is also generally called a metric. Contrary to Palatini approach, in this case the fundamental degree of freedom is no longer a connection but a new metric instead.

Using these techniques  we provide a full non-perturbative proof that the action in eq.~(\ref{initial}) contains propagating ghost degrees of freedom. This was already indicated  in ref. \cite{previous} using a perturbative analysis.
Moreover, we were able to show that contrary to our expectations and of those expressed in ref. \cite{previous} adding a standard Einstein-Hilbert term with an arbitrary coefficient to eq. \eqref{initial}
introduces new unhealthy degrees of freedom. Non-linear generalisations of such models also contain ghosts. Hence, we are lead to believe that a stable gravity model within the framework of bimetric variational formalism does not exist.

However, it is possible to construct viable models along similar lines of thought. We show this by considering the connection as a variation of the Levi-Civita version. Several simplest cases of this kind are demonstrated as examples in section~\ref{secHealthy}. We find variation with a scalar field particularly interesting. That  can give rise to a k-essence type  action and potentially be useful in building inflationary models or explaining  coincidence problems of accelerated expansion of the late Universe.

~

{\bf Acknowledgements.} AG is supported by the Saint Petersburg State University research grant No. 11.38.660.2013 and by Dynasty Foundation and was supported by the Russian Foundation for Basic Research grant No. 12-02-31214 in the initial stages of the project. AG is grateful to the University of Helsinki  and the members of cosmology group for hospitality during his visit. MK is supported by the Academy of Finland grant 1263714. HJN is supported by Magnus Ehrnrooth foundation and Vilho, Yrj\"o and Kalle V\"ais\"al\"a Foundation.

\bibliographystyle{unsrtnat}
\bibliography{bimetricRefs}

\end{document}